\documentstyle[floats,prl,aps]{revtex}

\input{epsf}

\begin{document}

\preprint{FERMILAB--Pub--97/002-A}
\title{CDM Models with a Smooth Component}
\author{Michael S. Turner$^{1,2,3}$ and Martin White$^{2,3}$}
\address{$^1$Department of Physics, The University of Chicago,
Chicago, IL~~60637-1433\\
$^2$NASA/Fermilab Astrophysics Center,
Fermi National Accelerator Laboratory,
Batavia, IL~~60510-0500\\
$^3$Department of Astronomy \& Astrophysics, Enrico Fermi Institute\\
The University of Chicago, Chicago, IL~~60637-1433}

\twocolumn[
\maketitle
\widetext

\begin{abstract}
The inflationary prediction of a flat Universe is at odds with
current determinations of the matter density ($\Omega_M\simeq0.2-0.4$).
This dilemma can be resolved if a smooth component contributes the remaining
energy density ($\Omega_X=1-\Omega_M$).
We parameterize the smooth component by its equation of state, $p_X=w\rho_X$,
and show that $x$CDM with $w\simeq-0.6$, $\Omega_M\simeq0.3$ and $h\simeq0.7$
is the best fit to all present cosmological data.   Together, the position
of the peak in the CMB angular power spectrum and the Type Ia supernova
magnitude-redshift diagram provide a crucial test of $x$CDM.
\end{abstract}
\hskip 0.5truecm
\pacs{XX.XX.Es}
]
\narrowtext

{\em Introduction.}
Inflation is a bold and expansive cosmological paradigm which makes three
firm and testable predictions:  flat Universe; nearly scale-invariant spectrum
of density perturbations; and nearly scale-invariant spectrum of gravitational
waves \cite{3predictions}.
(The first prediction can be relaxed at the expense of more complicated models
and tuning the amount of inflation \cite{loomega}.)
Flatness implies that the total energy density is equal to the critical
density ($\Omega_{\rm TOT} =1$).  However, it makes no prediction about
the form(s) that the critical energy takes.

Together, the first and second predictions lead to the cold dark matter (CDM)
scenario of structure formation which holds that most of the matter consists
of slowly moving elementary particles such as axions or neutralinos and that
structure in the Universe developed hierarchically, from galaxies to clusters
of galaxies to superclusters.
Both the density perturbations and the gravitational waves lead to
characteristic signatures in the anisotropy of the Cosmic Microwave
Background Radiation (CMB) \cite{cmb}.

The CDM picture is generally consistent with a wide array of cosmological
observations:
CMB anisotropy, determinations of the power spectrum of inhomogeneity from
redshift surveys and peculiar-velocity measurements, the evolution of galaxies
as recently revealed by the Hubble Space Telescope and the Keck telescope,
x-ray studies of clusters of galaxies and more.
Actually, there are several CDM models, distinguished by their ``invisible''
matter content (e.g.~Ref.~\cite{dgt} and references therein):
baryons + CDM only (sCDM, s for simple);
baryons + CDM + neutrinos with $\Omega_\nu \sim 0.15$ ($\nu$CDM);
baryons + CDM + cosmological constant ($\Lambda$CDM);
baryons + CDM + larger energy density in relativistic particles ($\tau$CDM).
Cosmological parameters also affect the predictions of each model:
Hubble parameter $H_0=100h\,{\rm km\,s^{-1}\,{\rm Mpc}^{-1}}$,
baryon density $\Omega_{\rm B}h^2$, power-law index characterizing the
spectrum of density perturbations $n$, and gravitational radiation described
by the its contribution to the quadrupole CMB anisotropy relative to that of
density perturbations ($T/S$) and the power-law index characterizing its
spectrum ($n_T$).
For each CDM variant there are values of the cosmological parameters for which
the model is consistent with most -- but possibly not all -- of the data.

{\em Flatness problem.}
{}From the very beginning, the prediction of a flat Universe has been
troublesome:  Put simply there has never been strong evidence for $\Omega_M =1$.
Today, almost all determinations of the matter density are consistent
with $\Omega_M=0.2-0.4$ \cite{omegasummary}.
(This does provide general support for the existence of CDM since big-bang
nucleosynthesis (BBN) constrains
$0.007h^{-2}\le\Omega_{\rm B}\le 0.024h^{-2}<0.1$ for $h>0.5$ \cite{cst}.)
Strong support for $\Omega_M\sim0.3$ comes from measurements of peculiar
velocities and the cluster baryon fraction.
Relating galactic peculiar velocities to the distribution of galaxies allows
the mean density to be sampled in a very-large volume, about
$(30h^{-1}{\rm Mpc})^3$, and several studies indicate that $\Omega_M$
is at least 0.25, but probably significantly less than 1 \cite{pv}.
X-ray observations of clusters of galaxies determine the baryon-to-total mass
ratio in a system of sufficient size to be representative of the universal
value ($\Omega_{\rm B}/\Omega_M$).
This, together with the BBN value for $\Omega_{\rm B}$,
implies $\Omega_M(h/0.7)^{1/2}=(0.3\pm 0.2)$ \cite{cbf}.
Indirect support for $\Omega_M<1$ comes from the fact that a flat,
matter-dominated universe (age $t_0={2\over 3}H_0^{-1}$) may be too
young to be consistent with determinations of the age of the oldest
stars ($t_0=15\pm 2\,$Gyr) \cite{age} and the Hubble parameter
($h=0.7\pm 0.1$) \cite{hubbleconst}.

In defense of a flat, matter-dominated Universe it should be said that there
has yet to be a convincing measurement of the matter density in a sufficiently
large volume to provide a definitive determination of $\Omega_M$ -- important
systematic and interpretational uncertainties remain even in the
peculiar-velocity and cluster-baryon-fraction methods.
While the age of the Universe coupled with large values of
the Hubble parameter argue for $\Omega_M <1$, the errors in $t_0$
and $h$ are still significant.   Finally, some methods continue to favor
higher values of $\Omega_M$: velocity power-spectrum measurements,
redshift-space distortions, void outflow, linear vs.~non-linear
power-spectrum measurements, galaxy counts and the problem of galaxy
anti-biasing (see e.g.~Ref.~\cite{WVLS}).

\begin{figure}
\centerline{\epsfysize=7cm \epsfbox{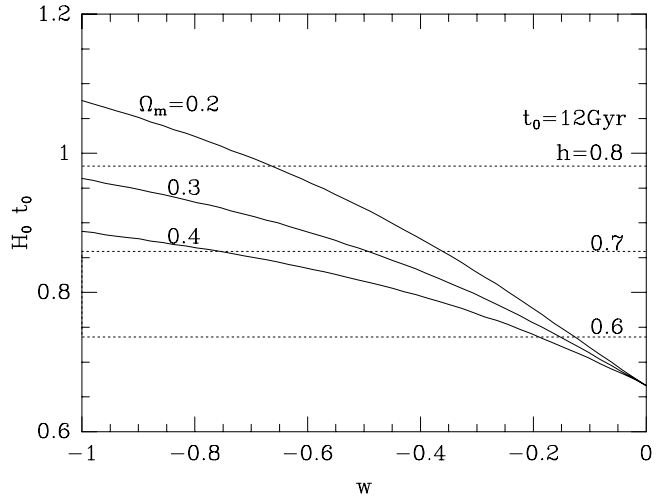}}
\caption{The age in Hubble units as a function of $w$ for $\Omega_M = 0.2$,
0.3 and 0.4.  Horizontal lines indicate the value of $H_0t_0$ required for
$t_0=12\,$Gyr with $h=0.6$, 0.7 and 0.8.  Note that the ``age constraint''
is strongly $w$ dependent.}
\label{fig:H0t0}
\end{figure}

A cosmological constant can resolve the flatness dilemma \cite{tsk,pjep}.
Since it corresponds to a uniform energy density (vacuum energy) that does
not clump, its presence is not detected in determinations of the matter
density.  Because of the accelerated expansion associated with a cosmological
constant, the expansion age is larger for a given Hubble parameter
(see Fig.~\ref{fig:H0t0}).
Until very recently, $\Lambda$CDM was the model preferred by the
observations \cite{lcdmbestfit}.

Two problems now loom for $\Lambda$CDM:  The limits to $\Omega_\Lambda$ from
(1) the frequency of gravitational lensing of distant QSOs,
$\Omega_\Lambda<0.66$(95\%CL) \cite{kochanek}, and
(2) the magnitude-redshift (Hubble) diagram of Type Ia supernovae (SNe-Ia)
$\Omega_\Lambda<0.51$ (95\%) \cite{perlmutter}.
Neither deals a death blow to $\Lambda$CDM --
$\Omega_\Lambda$ as low as 0.5 still retains many of the beneficial features
and several systematic uncertainties associated with the SNe-Ia determination
remain -- but a dark shadow has been cast.

{\em $x$CDM.}
Though $\Lambda$CDM is the ``best fit'' CDM model, the theoretical motivation
is weak.  The best argument for considering the tiny vacuum energy required,
$\rho_{\rm VAC} \sim 10^{-8}{\rm eV}^4$, is the absence of a reliable
calculation of the quantum vacuum energy \cite{weinberg}.
(Naive estimates of the vacuum energy range from 50 to 125 orders of magnitude
larger than this!).
Given the weak motivation for a cosmological constant and the apparent
observational evidence against one, as well as the strong motivation for
inflation and the evidence against $\Omega_M=1$, we think it worthwhile
to take a broader view.

Other possibilities have been suggested for a smooth component \cite{ct}:
relativistic particles \cite{tsk}; a tangled network of light
strings \cite{vilenkin}; texture \cite{davis};
and a decaying cosmological constant (i.e., scalar-field
energy) \cite{decaycc,sf1,sf2}.
For definiteness, as well as to facilitate a comprehensive analysis, we
parameterize the effective equation of state of the unknown, smooth component
by $w\equiv p_X/\rho_X$ with $w<0$ \cite{pjspu250}.
The energy density of the smooth component $\rho_X$ decreases as
$R^{-3(1+w)}$ where $R(t)$ is the cosmic scale factor;
vacuum energy corresponds to $w=-1$ and texture or tangled
strings correspond to $w=-{1\over 3}$.

For the reasons described above, we insist that X matter remain approximately
smooth on all scales.  Naively a component with $w<0$ should be highly
unstable to the growth of small-scale perturbations.  However, vacuum energy, by
definition, is constant in space and time.  Tangled strings, relativistic
particles, and scalar-field energy are all relativistic by nature and
hence very ``stiff;'' thus, in spite of the clumping of matter around
them, they should remain (nearly) smooth.
(In fact, it has been shown \cite{sf1,sf2} that scalar-field energy
remains approximately smooth.)
Relativistic particles, by virtue of their high speeds, do not clump
\cite{tsk}.
Likewise, it is easy to show that the effect of clumpy matter on an otherwise
straight string segment is to bend it (similar to the bending of light) by
an angle of order $\delta\Phi/c^2$, where $\delta\Phi\sim{\cal O}(10^{-5})$
is the typical magnitude of the large-scale perturbed gravitational potential
in the Universe.
Thus, a tangled string network should remain approximately smooth.
We consider our smoothness (or stiff X component) approximation to be a
reasonable starting point \cite{lateisw}.

\begin{figure}
\centerline{\epsfysize=7cm \epsfbox{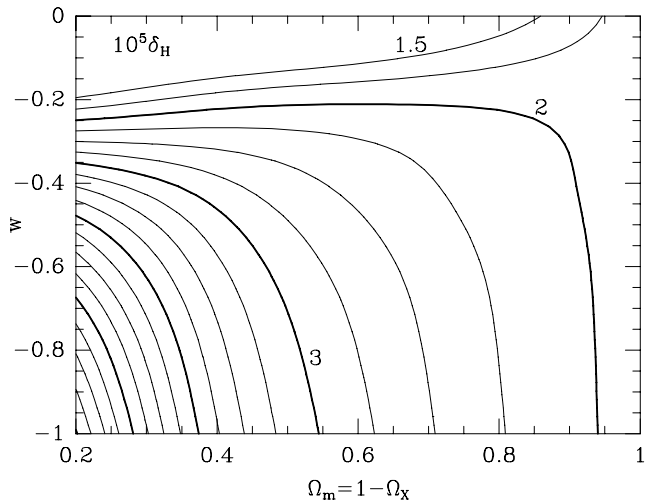}}
\caption{Power-spectrum normalization $10^5\delta_H$ as determined
by the four-year {\sl COBE\/} DMR results.  The {\sl COBE\/} $1\sigma$
error is approximately 10\%.}
\label{fig:delh}
\end{figure}

We note that there are reasons for only considering $w<0$.
The first is the age problem, which is even more severe for $w\ge 0$
(see Fig.~\ref{fig:H0t0}).
The second is that for $w>0$ the energy density in the smooth component
decreases faster than $R^{-3}$, implying that the ratio of the energy
density in the smooth component to the matter component was even larger at
earlier times.  This suppresses the growth of density perturbations,
and when the spectrum of density perturbations is fixed on large scales by
{\sl COBE\/}, this leads to too little inhomogeneity on small scales
(see Fig.~\ref{fig:delh}) \cite{caveat1}.
The case $w=0$ corresponds to the smooth component behaving like
pressureless matter; if the smooth component clumped, but only on large
enough scales to evade detection ($>50h^{-1}$Mpc), the flatness problem could
be solved and the {\sl COBE\/} normalization would be the same as sCDM because
the growth of density perturbations on large scales would be unaffected.
However, the growth of perturbations on small scales would be affected and
the problem of producing sufficient small scale structure would be similar
to that of hot dark matter.  Thus, we dismiss this possibility.

The formation of cosmic structure in a CDM model is dictated by the power
spectrum of density perturbations.  There are two important changes brought
about by the presence of a smooth component:  the normalization of the
power spectrum based upon the accurate {\sl COBE\/} determination of CMB
anisotropy on angular scales of around $10^\circ$ and the transfer function
that describes the growth of density perturbations from the inflationary epoch
to the present.
For fixed inflationary perturbations, CMB anisotropy on {\sl COBE\/}
scales is larger (due to the integrated Sachs-Wolfe effect \cite{ISW});
because of the smooth component there is less growth of density
perturbations from the inflationary period until the present.

We use the {\sl COBE\/} four-year results \cite{cobe4yr} to normalize the
power spectrum (assuming negligible gravity waves and $n=1$) using
the method of Ref.~\cite{BunWhi}.
Writing the (linear) power spectrum today as
\begin{eqnarray}
P(k) & \equiv & \langle |\delta_k|^2\rangle = {2\pi^2\over H_0^3}
        \delta_H^2 (k/H_0)^n T^2(k) \\
T(q) & = & \frac{\ln \left(1+2.34q \right)}{2.34q} \times
        \\ \nonumber
& & \hspace*{-0.5cm}
        \left[1+3.89q+(16.1q)^2+(5.46q)^3+(6.71q)^4\right]^{-1/4} \,,
\end{eqnarray}
where $\delta_k$ is the Fourier transform of the density field,
$q = k/h\Gamma$ and $\Gamma\simeq \Omega_Mh$ is the ``shape'' parameter
\cite{BBKS,Gamma}. The quantity $\delta_H$, which corresponds to the amplitude
of density perturbations on the Hubble scale today, is a convenient
normalization whose value is shown as a function of $\Omega_X$ and $w$
in Fig.~\ref{fig:delh}.
The transfer function, $T(k)$, is well fit by the form quoted for
$w<0$, with more small-scale power than this form predicts when $w\to0$.

\begin{figure}
\centerline{\epsfysize=7cm \epsfbox{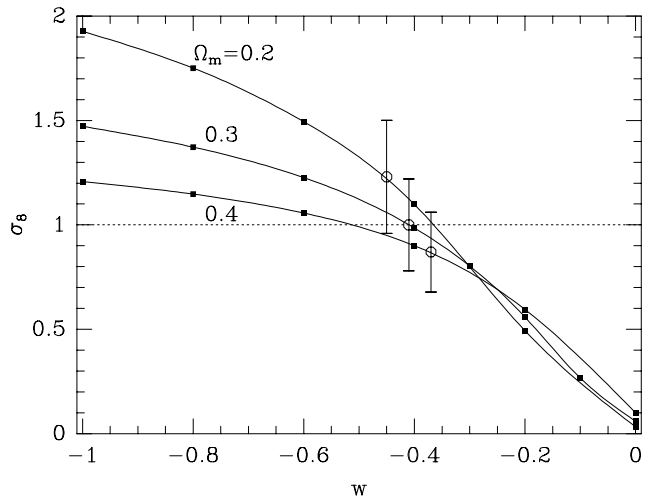}}
\caption{$\sigma_8$ as a function of $w$ with $\Gamma=0.25$ and $n=1$.
Points with $2\sigma$ error bars on the curves show $\sigma_8$
from cluster abundance for that value of $\Omega_M$.
Note, $\sigma_8$ scales as $(\Gamma/0.25)^{1.3}\exp[-3.1(n-1)]$.}
\label{fig:sig8}
\end{figure}

There are many constraints on CDM models.  The two most stringent for the
power spectrum in low-$\Omega_M$ models are: the shape parameter
$\Gamma = 0.25\pm0.05$ (for $n=1$) \cite{shape} and the abundance of
rich clusters.
The latter can be reduced to a constraint on $\sigma_8$ (the {\em rms}
mass fluctuation in spheres of radius $8h^{-1}$Mpc),
\begin{equation}
\sigma_r^2 \equiv \int_0^\infty {dk\over k}\ {k^3P(k)\over 2\pi^2}\,
 \left( {3j_1(kr)\over kr}\right)^2
\end{equation}
with $r=8\,h^{-1}$Mpc.  There is no consensus on the precise
value of $\sigma_8$ or its scaling with $\Omega_M$; differences
arise due to different input data and calculational schemes \cite{sigma8}.
Further, the scaling with $\Omega_M$ depends slightly upon $w$,
through the relation between virial mass and cluster temperature.
Nevertheless, there is a general consensus about this important
constraint and as a middle-of-the-road estimate we use
$\sigma_8=(0.55\pm0.06)\Omega_M^{-0.5}$ which is consistent with most
published estimates \cite{sigma8} and slightly conservative (low $\sigma_8$)
near $\Omega_M\sim0.3$.

There are two nice features of $x$CDM:
The shape constraint can be satisfied with $h\sim0.7$ and $\Omega_X\sim 0.6$
for which the $\sigma_8$ constraint can be readily satisfied with $w\sim-1/2$
(see Fig.~\ref{fig:sig8}).
For $\Lambda$CDM ($w=-1$) tilt (i.e.~$n<1$) and/or gravity waves are needed
to reduce $\sigma_8$ and for an open Universe (closely approximated by
$w=-{1\over 3}$) $\sigma_8$ is too small unless $\Omega_M$ is large
or $n>1$ \cite{whitesilk}.

\begin{figure}
\centerline{\epsfysize=7cm \epsfbox{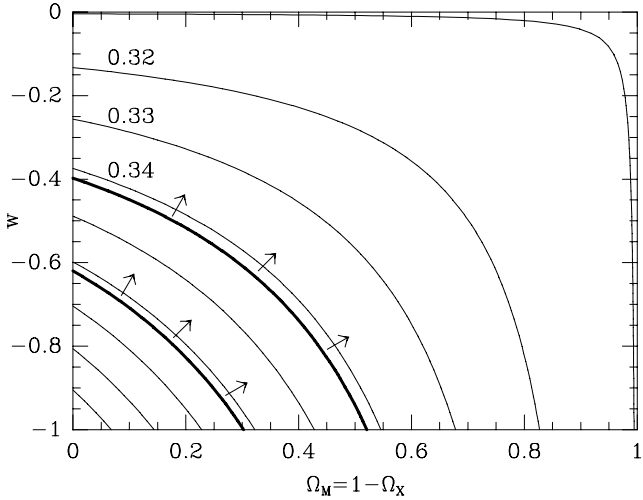}}
\caption{Constraints from the SNe-Ia magnitude-redshift diagram.
Contours are of $H_0r(z=0.4)$ from 0.31 to 0.39 in
steps of 0.01.  The thick contours are the current 95\%CL and
68\%CL limits, and arrows indicate that values to the upper right
of these curves are favored.  (For $w=-1$ the constraint shown here is
less stringent than that in Ref.~\protect\cite{perlmutter} because we have
two free parameters rather than one.)}
\label{fig:rz}
\end{figure}

Next, we turn to the two worries of $\Lambda$CDM --
the frequency of QSO lensing and the SNe-Ia constraint.
Both involve the increased distance to a given redshift that comes with
$\Lambda$.  The proper distance today is given by the Robertson-Walker
radial coordinate \cite{pade}
\begin{eqnarray}
r(z) & = & \int_0^z {dz\over H(z)} =
  H_0\left[ z - {1\over 2}(1+q_0)z^2 + \cdots \right], \\
H^2(z) & = & H_0^2\left[ (1+z)^3\Omega_M + (1+z)^{3(1+w)}\Omega_X \right],
\end{eqnarray}
and the deceleration parameter
$q_0=-{\ddot R}_0/R_0H_0^2={1\over 2}+{3\over2}w\Omega_X$.  Note, $r(z)$
increases with decreasing $w$; this leads to more volume and more lenses
between us and a QSO at redshift $z$ and a higher frequency of lensing.

While the SNe-Ia limits on the distance redshift relation \cite{perlmutter}
are quoted for a flat universe with cosmological constant,
they are readily translated into a constraint on $r$.
Since the seven distant SNe-Ia have redshifts $z\sim 0.4$, that constraint
can be expressed as $0.287 (0.271)<H_0r(z=0.4)<0.342 (0.362)$ at
68\%CL (95\%CL) \cite{perlmutter2}.
Their results constrain $\Omega_M$ and $w$ (see Fig.~\ref{fig:rz}).
Soon, Perlmutter's group should have results based on nearly four times as
many SNe-Ia's and another group (The High-z Supernova Team)
should have results based on a comparable number of SNe-Ia's.
This will sharpen this important constraint to $w$ significantly.

\begin{figure}
\centerline{\epsfysize=7cm \epsfbox{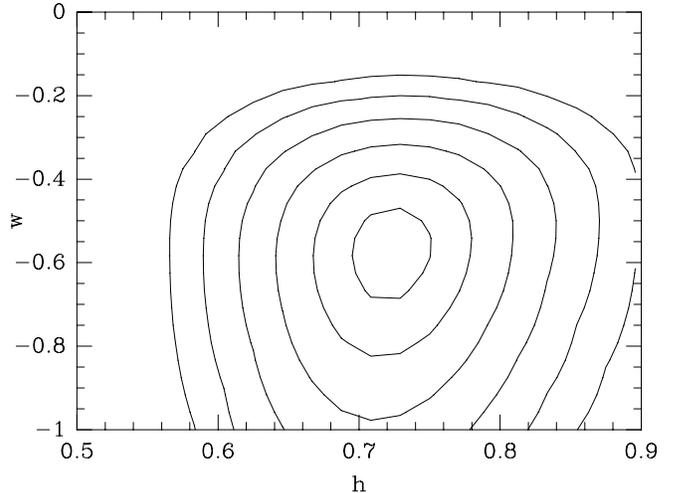}}
\caption{The likelihood, marginalized over $\Omega_M$.  Contours are in
units of ${1\over2}\sigma$.  (See text for the cosmological data
and conservative error bars used.)}
\label{fig:lik}
\end{figure}

{\em Concluding remarks.}
Inflation is a bold and compelling idea.  It predicts a flat Universe, but
not the form which the critical energy density takes.
Because of increasing evidence that the matter density is significantly less
than the critical density, as well as the attractiveness of inflation and
the successes of CDM, we have explored the possibility that most of the
critical energy density resides in a smooth component of unknown nature,
with equation of state $p_X=w\rho_X$ ($w<0$).
Increasing $w$ to around $-0.6$ retains the attractive features of
$\Lambda$CDM and resolves the conflict with the SNe-Ia constraint; further
tilt and/or gravity waves are not required to obtain the correct number of
rich clusters observed at present.

For the sake of illustration we have used the following cosmological
data to find the best fit $x$CDM model:
$t_0=15\pm 2\,$Gyr, $h=0.7\pm 0.07$, $\Omega_{\rm B}h^2=0.02$,
$\Gamma=0.25\pm 0.1$, $\sigma_8\Omega_M^{0.5}=0.55\pm 0.1$,
$[H_0r(z=0.4)]^2=0.10\pm 0.015$ and the {\sl COBE\/} four-year data set.
(For several constraints we have inflated the error bars to be conservative.)
We have marginalized over $\Omega_M$ with prior $\Omega_M=0.3\pm 0.05$.
For $n=1$, an $x$CDM model with $w=-0.6$ and $h=0.7$ has maximum likelihood
(see Fig.~\ref{fig:lik}).  (The unmarginalized likelihood prefers
$\Omega_M\sim0.4$, $h=0.7$ and $w=-0.4$ but is quite broad.)
In passing, we note that a ``tangled'' network of walls or a wall wrapped
around the Universe (supposing space is $S^2\times S^1$) would lead to a
smooth component with $w=-{2\over 3}$.

Introducing $w$ to the list of CDM parameters brings the total to at least
ten ($w$, $n$, $h$, $\Omega_{\rm B}h^2$, $n_T$, $T/S$, $\Omega_{\rm TOT}$,
$\Omega_\nu$, $\Omega_X$, and $\Omega_{\rm RAD}$).
While this is a daunting number, the flood of cosmological data coming --
larger redshift surveys, accurate measurements of the expansion rate and
deceleration rate of the Universe, high resolution observations of clusters
with X-rays, the Sunyaev-Zel'dovich effect and weak lensing, studies of
galactic evolution by HST and Keck, and especially measurements of CMB
anisotropy on angular scales from arcminutes to tens of degrees --
should eventually overdetermine the parameters of CDM + inflation.
Then the data will not only sharply test inflation, but also discriminate
between different CDM models and even provide information about the underlying
inflationary potential \cite{reconstruct}.

In the near term, the SNe-Ia magnitude-redshift diagram and CMB angular power
spectrum will provide an important test of $x$CDM:
the position of features (e.g., the first peak) in the angular power spectrum
tests flatness, but is less sensitive to $w$, and given $\Omega_{\rm TOT}$
(and $\Omega_M$), SNe-Ia can determine $w$ (see Fig.~\ref{fig:rz}).

\noindent{\em Acknowledgments.}
This work was supported by the DoE (at Chicago and Fermilab) and by
the NASA (at Fermilab by grant NAG 5-2788).


\begin{thebibliography}{99}
\bibitem{lateisw}
In addition to the clumping that is caused by the lumpy matter distribution
in the Universe, the X component could also have been slightly perturbed
(density variations of order $10^{-5}$) to begin with; this would naturally
occur if the underlying density perturbations are curvature perturbations
(as predicted by inflation).
The small amount of clumping of scalar-field energy, tangled strings, or any
approximately smooth component, while certainly unimportant for considerations
of the distribution of matter today or the evolution of the cosmic-scale
factor, will have a small effect on large-angle CMB anisotropy and hence
{\sl COBE\/} normalization.
When the X component is becoming dominant (at relatively recent times),
it affects large-angle CMB anisotropy due to the spatially and time-varying
gravitational potential it creates (``late-time, integrated Sachs-Wolfe
effect'').  While the effect of the time-varying potential is included
in our work, and is significant, the effect of spatial variations is not.
However, in Ref.~\cite{sf1} the effect of spatial variations
on the {\sl COBE\/} normalization is shown to be small for
scalar-field energy.
\bibitem{caveat1}
The second problem could by circumvented by having the smooth component come
into existence recently, e.g., a relativistic sea of particles ($w=1/3$) can
be created by the decay of unstable CDM (see Ref. \cite{tsk}).
\bibitem{pade}
The Taylor series expansion is not very accurate, even for $z\sim 0.4$.
Pad\'{e} approximants provide a more accurate approximation; e.g.,
the second-order Pad\'{e} approximant is $H_0r(z) = z/[1+(1+q_0)z/2]$.
For all of our results we have numerically integrated the expression for
$r(z)$.
\bibitem{perlmutter2}
The 95\% cl limit to $\Omega_\Lambda$ quoted in Ref. \cite{perlmutter}
is more stringent than that which follows from the 95\% cl limit to $r(z)$
because the authors of Ref. \cite{perlmutter} reduce the number of free
parameters from two ($\Omega_M$, $\Omega_\Lambda$) to one.
\end{thebibliography}
\end{document}